# Extreme Enhancement-Mode Operation Accumulation Channel Hydrogen-Terminated Diamond FETs with $V_{th}$ < -6V and High On-Current

*Chunlin Qu[a,*], Isha Maini[a,1], Qing Guo[a,2], Alastair Stacey[b,c,3], David A. J. Moran[a,4]*

[a] University of Glasgow, James Watt School of Engineering, Glasgow, UK

[b] School of Science, RMIT University, Melbourne, Victoria 3010, Australia

[c] Princeton Plasma Physics Laboratory, Princeton University, Princeton, NJ, 08540 USA

[*] Corresponding Author Email: chunlin.qu@glasgow.ac.uk

[1] isha.maini@glasgow.ac.uk

[2] q.guo.1@research.gla.ac.uk

[3] alastair.stacey@rmit.edu.au

[4] david.moran@glasgow.ac.uk

Data Availability statement

*The data that support the findings of this study are available from the corresponding author upon reasonable request.*

Funding Statement

*Funding information is not available.*

Conflict of interest

*The authors declare that they have no known competing financial interests or personal relationships that could have appeared to influence the work reported in this paper.*

Abstract

In this work we demonstrate a new Field Effect Transistor device concept based on hydrogen-terminated diamond (H-diamond) that operates in an Accumulation Channel rather than Transfer Doping regime. Our FET devices demonstrate both extreme enhancement-mode operation and high on-current with improved channel charge mobility compared to Transfer-Doped equivalents. Electron-beam evaporated Al$_2$O$_3$ is used on H-diamond to suppress the Transfer Doping mechanism and produce an extremely high ungated channel resistance. A high-quality H-diamond surface with an unpinned Fermi level is crucially achieved, allowing for formation of a high-density hole accumulation layer by gating the entire device channel which is encapsulated in dual-stacks of Al$_2$O$_3$. Completed devices with gate/channel length of 1 μm demonstrate record threshold voltage < -6 V with on-current > 80 mA/mm. Carrier density and mobility figures extracted by CV analysis indicate high 2D charge density of ~ $2 \times 10^{12}\ cm^{-2}$ and increased hole mobility of $110\ cm^2/V \cdot s$ in comparison with more traditional Transfer-Doped H-diamond FETs. These results demonstrate the most negative threshold voltage yet reported for H-diamond FETs and highlight a powerful new strategy to greatly improve carrier mobility and enable enhanced high power and high frequency diamond transistor performance.

1. Introduction

Enhancement Mode (E-mode) or "Normally Off" operation is an essential requirement for many modern solid-state electronic devices such as Field Effect Transistors (FETs), where extremely high resistance in their default (unpowered) off-state is an essential requirement to meet rigorous safety standards. This is particularly true for Wide Bandgap (WBG) and Ultra-Wide Bandgap (UWBG) semiconductor technologies given their predominant focus on high-power electronic applications. Conversely, such devices must also demonstrate sufficiently high on-state currents and low on-resistance to reduce associated electrical losses and be competitively energy efficient. Simultaneously achieving both E-mode operation with high on-current can be extremely challenging for lateral WBG/UWBG devices such as FETs, as evidenced by the significant research effort and associated differing strategies applied in E-mode GaN FET technology [1,2]. To attain E-mode operation in such a scenario requires suppression of the carrier density in the channel, by engineering the gate interface in a way which makes it difficult to achieve any considerable carrier concentration without the electrostatic contribution of a large gate-bias. The challenge herein is to sufficiently deplete the channel in the absence of a gate bias, while ensuring the Fermi level at the gate interface remains sufficiently unpinned to allow for accumulation of a high carrier concentration with increasing gate-field to induce high on-current when the device is turned on.

Diamond is a UWBG semiconductor with significant potential for high-power electronic applications and is therefore subject to similar device requirements in terms of efficiency and high-performance E-mode operation. Challenges presented by large activation energy dopant species in substitutionally-doped diamond however have limited development of high performance devices. Recent developments in Transfer Doping of hydrogen-terminated diamond (H-diamond) as an alternative to substitutional doping have led to significant progress in electronic device performance [3]. For example, *p*-type Transfer-Doped diamond-based FET technology has demonstrated impressive on-current values up to 1.3 A/mm [4], RF power operation of 4.2 W/mm at 2 GHz [5], cut-off frequencies up to 70 GHz [6] and off-state breakdown up to 2 kV [7]. Transfer Doping overcomes the limitations of substitutional doping by forming a high-density (typically $10^{12}$ cm$^{-2}$ to $10^{13}$ cm$^{-2}$) 2D sub-surface hole channel within H-diamond in the presence of a suitable surface electron acceptor material, without the requirement to introduce impurity atoms into the diamond lattice [8,9]. Some successes have been reported in the demonstration of E-mode Transfer-Doped diamond FETs, including typical threshold voltage values > -3 V [7,10–17]. The high carrier densities that result from

Transfer Doping make it difficult to achieve significantly negative $V_{th}$ values and hence E-mode operation without damaging or suppressing the Transfer Doping mechanism however, thus resulting in reduced on-current values compared to equivalent Transfer-Doped depletion mode (D-mode) diamond devices [18,19]. Another drawback of Transfer Doping is the significant reduction in hole mobility (~ 70 $cm^2/V \cdot s$) [9,20] in comparison with the high values achievable in intrinsic, undoped diamond (~2000 $cm^2/V \cdot s$) [21]. This mobility degradation is predominantly attributed to Coulomb scattering from the close-proximity electrons transferred to and trapped within the surface acceptor material that forms an intrinsic part of the Transfer Doping process [22]. Thus, there appears to exist an intimate coupling and trade-off between the 2D hole density and hole mobility in Transfer-Doped H-diamond.

In this work we demonstrate a new strategy to exploit the positive attributes of the Fermi-unpinned H-diamond surface to overcome the limitations of Transfer Doping by instead establishing an Accumulation Channel architecture (Fig. 1). Specifically, the Transfer Doping process is actively suppressed through deposition of an electron-beam evaporated $Al_2O_3$ dielectric layer onto H-diamond following an *in-situ* anneal of 350ºC to form a residue-free and high-quality Fermi-unpinned interface. The use of electron-beam evaporated $Al_2O_3$ here is in contrast to Atomic Layer Deposited (ALD) $Al_2O_3$, which conversely has been reported to induce high carrier density Transfer Doping in H-diamond, the mechanism for which is still under debate[23–27]. A high-density hole accumulation layer is then formed in the presented Accumulation-Channel (AC) FET device (Fig. 1 - left) only upon strongly biasing the gate, which encompasses the entire device channel. For comparison, a generic Transfer-Doped H-diamond FET structure is also presented in Fig. 1 (right), whereby a continuous hole channel is spontaneously formed by electron transfer from the diamond valance band to a surface electron acceptor material. Accumulation Channel devices we report here demonstrate the most negative threshold voltage yet reported for H-diamond FETs combined with high on-current and improved hole mobility in comparison with typical values reported for Transfer-Doped FETs. These results demonstrate a new approach to simultaneously deliver unparalleled normally-off device operation combined with enhanced hole mobility for advanced high power and RF performance in diamond transistor technology.

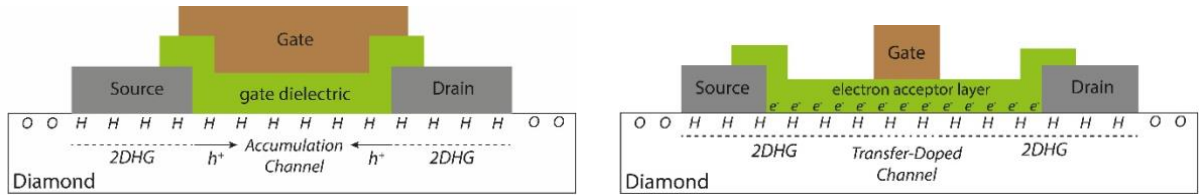

*Fig. 1 – Accumulation Channel H-diamond FET structure concept (left) in contrast with a "typical" Transfer-Doped H-Diamond FET structure (right).*

2. Materials and Methods

Single crystal {001} 4.5 x 4.5 mm diamond substrates purchased from Element Six were utilized in this work. Received substrates were scaif polished to achieve a uniform surface roughness of $R_q = 0.77$ nm (RMS) as verified by AFM. The substrate used in this study was then acid cleaned using the following two-stages process: firstly, it was immersed for 15 minutes in a boiling solution of $HNO_3$ (65%) and HCl (37%) in a 1:1 ratio, followed by the second stage clean wherein it was immersed in boiling $HNO_3$ (65%) and $H_2SO_4$ (95%) in a solution with a 1:3 ratio. Prior to hydrogen termination, the substrate was treated with 10 minutes UV ozone. Hydrogen termination was undertaken within a Seki 6500 reactor, following a short-growth and hydrogen plasma exposure method, under the following conditions: The substrate was subjected to plasma treatment and heated to approximately 800°C. It was then cooled under a gradually decreasing plasma intensity until the sample temperature reached approximately 700°C, at which point the plasma was turned off. Four-contact probe Van der Pauw (VDP) measurements of the hydrogen-terminated surface in air indicated a sheet resistance of ~ 50 k$\Omega$/□ due to air-induced Transfer Doping. Surface roughness was extracted by AFM after H-termination to be 1.04 nm (RMS). An 80 nm-thick Au layer was then deposited across the surface of the entire H-diamond substrate to serve as a protective layer during subsequent processing steps. To produce the Accumulation Channel FETs (Fig. 2) and associated test structures, the following fabrication steps were performed in the order described: 1. Alignment markers, 2. Isolation for device geometry definition, 3. Ohmic contact formation, 4. $Al_2O_3$ layer(s) deposition, 5. Contact pad formation, 6. Gate formation. Electron beam lithography using a Raith EPBG tool and PMMA resist was used for pattern definition and a 1:4 $KI/I_2$: $H_2O$, room-temperature etch solution was used to remove the Au-protective layer when required during each processing level. For device and test structure isolation, an oxygen plasma process was used to oxygen-terminate the exposed diamond surface following Au etch removal in select areas. Ohmic contacts were formed by electron-beam evaporation of 80 nm of Pd after selective Au etch in the ohmic contact regions, resulting in a source-drain separation of 1 µm. Pd was utilized as an ohmic contact metal due to its resilience to KI/I$_2$ etch chemistry and its comparable work function to Au. Following ohmic contact formation, KI/I$_2$ was used to remove all residual Au from the substrate surface. Deposition of the Al$_2$O$_3$ layers was then undertaken by a 2-stage process: Immediately prior to Al$_2$O$_3$ deposition, the substrate was treated with a 350 °C *in-situ* anneal for 30 minutes at a vacuum of $1 \times 10^{-7}\ mbar$, to thermally remove any residual atmospheric species from the surface[8,20]. 35 nm of Al$_2$O$_3$ was then deposited across the entire substrate by *electron beam* evaporation, without breaking vacuum after the 350 °C anneal. A second Al$_2$O$_3$ layer with a thickness of 15 nm was then deposited by conformal plasma-enhanced ALD across the entire substrate with the primary purpose of coating any potentially exposed ohmic contact metal following the initial electron-beam Al$_2$O$_3$ layer deposition to avoid physical contact with the subsequently deposited gate contact (Fig 2). To ensure resilience of the metal contacts to probing procedures during electrical characterization, contact pads comprising 20 nm Ti / 130 nm Au were defined and deposited with an overlap onto the Pd ohmic contacts. A 10:1 buffered oxide etchant (HF) was used to selectively remove both electron-beam and ALD Al$_2$O$_3$ layers in regions over the Pd contacts to ensure robust electrical connection to the contact pad metals. Finally, a 3 µm length gate contact comprising 20 nm Al / 20 nm Pt / 100 nm Au was formed on the surface ALD Al$_2$O$_3$ layer across the entire 1 µm long device channel and overlapped by 1 µm onto each of the Pd ohmic contacts. A cross-sectional schematic of the completed device layout is presented in Fig 2. and a top-down optical microscope image presented in Fig. 3.

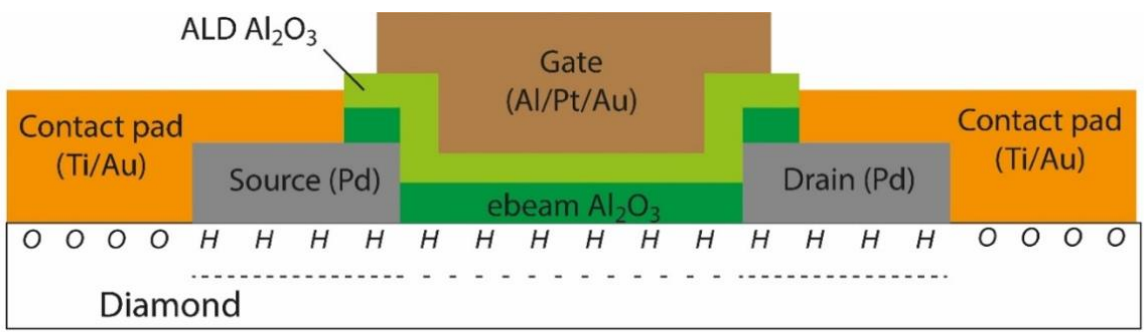

*Fig 2 - Schematic cross section of completed Accumulation Channel H-diamond FET with 1 µm gate/channel length and 25 µm width.*

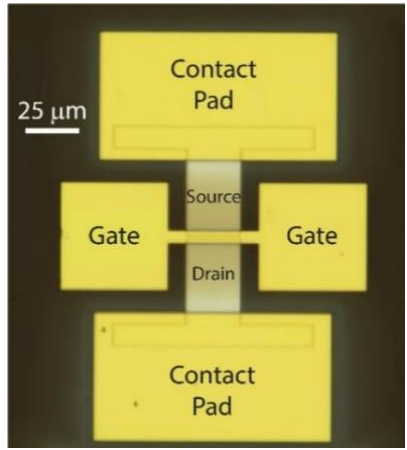

*Fig. 3 - Top-down optical microscope image of completed Accumulation Channel H-diamond FET with 1 μm gate/channel length and 25 μm width.*

3. Results

Typical IV Output ($I_d$ *vs* $V_{ds}$) characteristics extracted for a 1 $\mu$m gate/channel length, 25 $\mu$m wide Accumulation Channel FET are presented in Fig 4. Gate voltage ($V_{gs}$) was stepped in a range from -13 V to 0 V while source-drain bias ($V_{ds}$) was swept from 0 V to -10 V. A maximum on-current ($I_{on}$) ~80 mA/mm is achieved at a $V_{gs}$ of -13 V and $V_{ds}$ of -10 V. Significant suppression of the drain current is already observed by the point $V_{gs}$ is increased from -13 V to -7 V, indicating a strongly negative threshold voltage.

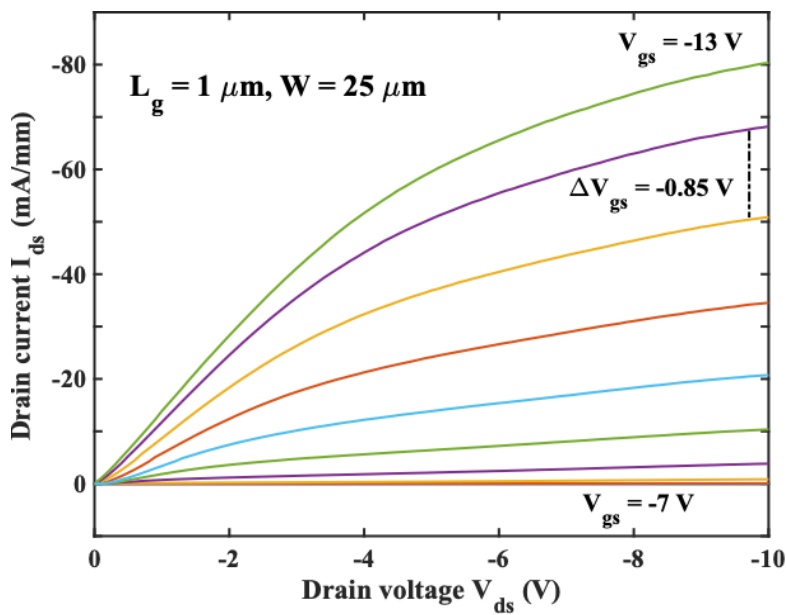

*Fig. 4. Output ($I_d$ vs $V_{ds}$ for stepped $V_{gs}$) IV characteristics for a typical 1 μm gate/channel length, 25 μm wide Accumulation Channel FET.*

To allow for detailed inspection and quantification of the threshold voltage, the logarithmic transfer ($I_d$ vs $V_{gs}$ for stepped $V_{ds}$) IV characteristics for this device are presented in Fig. 5.

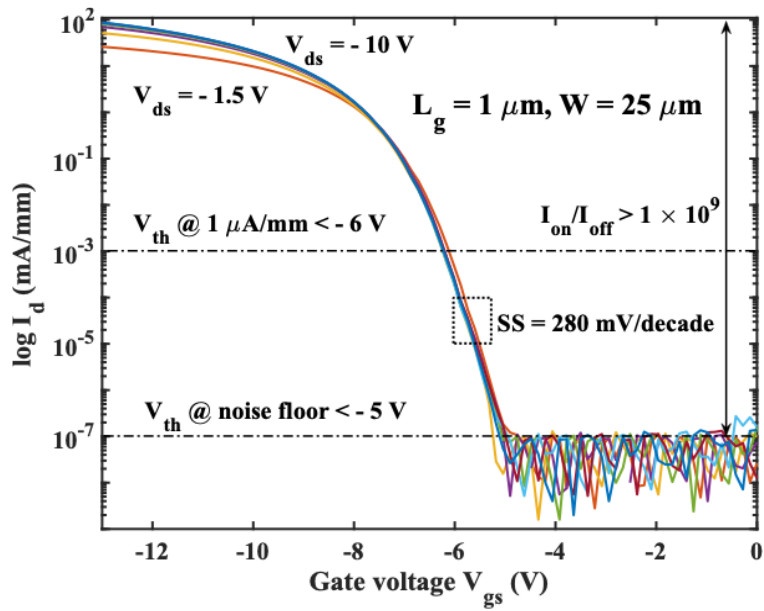

*Fig. 5. Logarithmic Transfer (log $I_d$ vs $V_{gs}$ for stepped $V_{ds}$) IV characteristics for a typical 1 μm gate/channel length, 25 μm wide Accumulation Channel FET.*

For a set off-state current density $I_{off}$ = 1 μA/mm, a threshold voltage, $V_{th}$ < - 6 V is extracted across the inspected $V_{ds}$ range of -1.5 V to – 10 V. Furthermore, only $I_d$ values below the noise floor of the measurement system were determined for $V_{gs}$ values larger than -5 V, demonstrating a very negative $V_{th}$ < -5 V for an extremely large $I_{on}/I_{off}$ ratio > 1 x $10^9$. A value of 280 mV/decade for Subthreshold Swing (SS) was determined for each value of $V_{ds}$ inspected. Gate leakage current in these devices also remained below the measurement capabilities of our measurement system (Keysight B1500A) across the entire inspected bias range (<40 nA/mm). Linear Transfer IV characteristics ($I_d$ & $g_m$ vs $V_{gs}$ for $V_{ds}$= -10 V) for this device are also presented in Fig. 6, demonstrating a peak extrinsic transconductance value ($g_m$) of 25 mS/mm.

Different techniques are often applied to extract FET threshold voltage, which may lead to varying reported values for similar technologies. To ensure thorough and honest benchmarking of the presented devices, other common $V_{th}$ extraction processes were also implemented for comparison: Through extrapolation from the presented linear $I_dV_{gs}$ response from the peak transconductance point[28] (Fig. 6)[28], an extremely negative threshold voltage value of – 9 V $V_{gs}$ is determined using this approach. Similarly, linear extrapolation of the root square Transfer characteristic (√$I_d$ vs $V_{gs}$ for $V_{ds}$ = -10 V)[29] shown in Fig. 7 also produces a more pronounced negative threshold value of -7 V $V_{gs}$ compared to the fixed off-state current value approach utilized in Fig. 5.

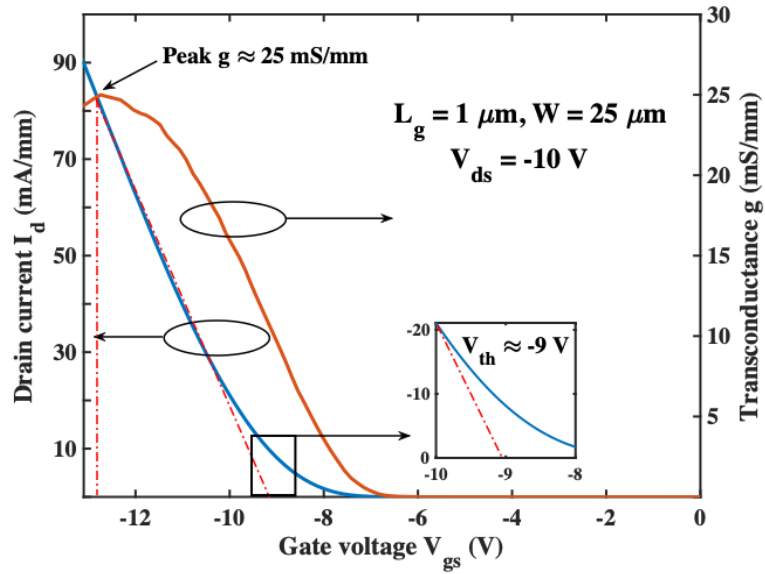

Fig. 6. Linear Transfer ($I_d$ and $g_m$ vs $V_{gs}$ for $V_{ds}$ = -10V) IV characteristics for a typical 1 μm gate/channel length, 25 μm wide Accumulation Channel FET.

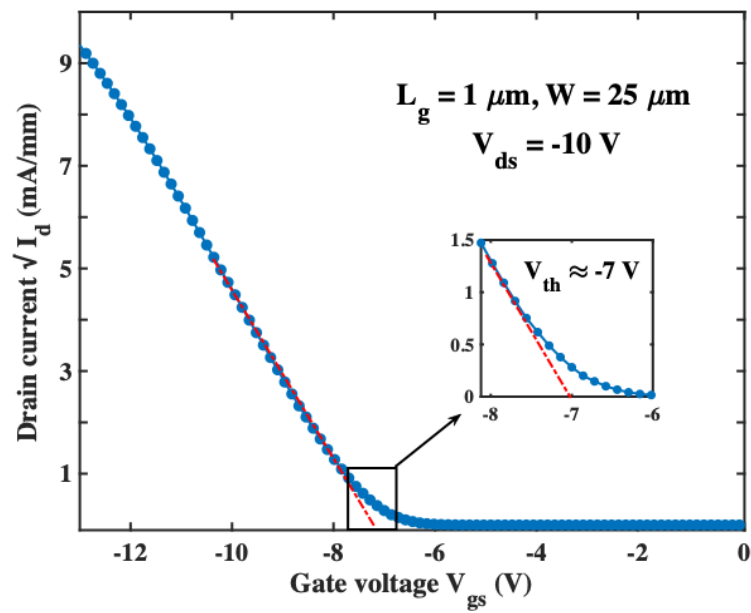

Fig. 7 Root square Transfer ($\sqrt{I_d}$ vs $V_{gs}$ for $V_{ds}$= -10 V) IV characteristics for a typical 1 μm gate/channel length, 25 μm wide Accumulation Channel FET.

The corresponding drain current at threshold voltage is found to be 6 mA/mm using the peak transconductance extraction method, 90 µA/mm using the root square drain current method and 1 µA/mm with fixed off-state current method. Thus, the fixed off-state current approach is prioritized in this work for device threshold voltage determination as this provides the most stringent extraction criteria as well as the highest $I_{on}/I_{off}$ ratio.

Off-state breakdown measurements as shown in Fig.8 were also performed on devices (at $V_{gs} = 0$ V), demonstrating a maximum breakdown voltage of ~ 23 V $V_{ds}$ and coinciding with a sharp increase in gate leakage current.

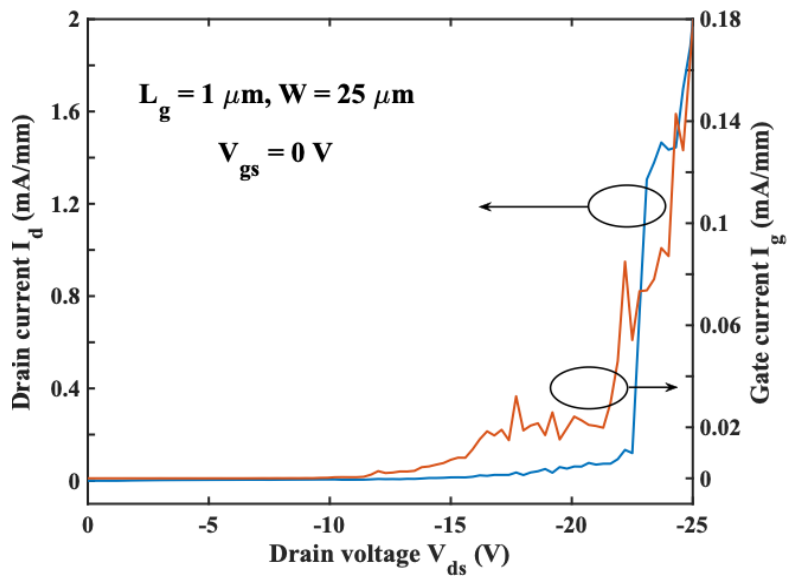

*Fig. 8. Off-state breakdown ($V_{gs} = 0$ V) measurement for a typical 1 µm gate/channel length, 25 µm wide Accumulation Channel FET.*

4. <u>Discussion</u>

The results presented in Figs. 4 to 7 demonstrate excellent on-state FET performance ($I_{on}$ approaching 100 mA/mm) combined with extremely low threshold voltage (less than -6 V). Despite the observed continuous increase in drain current with negative $V_{gs}$ within the reported gate bias range, irreversible degradation in device performance unfortunately was observed for $V_{gs}$ values lower than -13 V. This perhaps suggests high field-induced damage to the $Al_2O_3$ layers/interface with the H-diamond and thus demonstrates potential for access to even higher drain currents with better field management. Off-state breakdown voltage up to 23 V was extracted from devices, with breakdown occurring as a rapid increase in gate current, most likely associated with breakdown of the 15 nm thick ALD $Al_2O_3$ layer between the gate and ohmic contacts (Fig. 2). This is in good agreement with reported breakdown field values of ~ 10 to 30 MV/cm for 'thin' $Al_2O_3$ thin films [30,31]. Ohmic contact resistance, $R_c$, as extracted from Transmission Line Method (TLM) structures fabricated in parallel with FETs was 11.4 Ω·mm, which is larger than typical values reported for Transfer-Doped H-diamond (~ 5 Ω·mm [32,33]). This was most likely attributed to the significantly large starting sheet resistance of 50 kΩ/□ extracted for the H-diamond substrate prior to device fabrication, and the resultant low hole density within the channel beneath the Pd ohmic contacts. High on-currents are however still demonstrated with these devices, suggesting that further optimization of the ohmic contacts should result in even higher $I_{on}$ values. Modest values for peak extrinsic transconductance were also extracted (~25 mS/mm) in comparison with equivalent Transfer-Doped technology [19,34], again associated with the larger ohmic contact resistance but also the use of a relatively thick $Al_2O_3$ gate dielectric stack (~ 50 nm total thickness). These device results may also be compared with a previous study whereby we reported similar on-current values (~100 mA/mm) for Transfer-Doped H-diamond FETs utilizing electron-beam evaporated $Al_2O_3$ as a gate dielectric, albeit with a reduced gate length of 250 nm rather than 1 $\mu$m used here [35]. In contrast with the results reported here, a threshold voltage of ~ +1 V and hence depletion mode operation was achieved in these previously reported devices, demonstrating a difference of over 7 V in $V_{th}$ compared to devices presented herein that utilize the Accumulation Channel architecture. In this work, the marked decrease in threshold voltage is likely attributable to the inclusion of the 350 ℃ *in-situ* anneal prior to e-beam $Al_2O_3$ evaporation, which has been shown to remove residual surface adsorbates on the H-diamond surface [36,37], as is critical to the suppression of Transfer-Doping induced surface conductivity and consequent achievement of extreme enhancement-mode operation.

The elimination of the spontaneously formed hole channel was additionally experimentally verified through electrical characterization of ungated TLM and VDP structures that received the same annealing of the H-diamond surface at 350 ℃ and electron beam deposition of $Al_2O_3$ as implemented in FET devices. The resultant $R_{sh}$ was determined to be too large to be measurable by VDP, while TLM IV results demonstrated variable but very low current densities ~ 500 nA/mm. Furthermore, by maintaining the high-quality H-termination of the diamond surface, high device on-current is simultaneously achieved.

To better understand the low-field charge transport through the hole accumulation channel and associated dependency on gate bias, capacitance-voltage (CV) measurements were undertaken on devices. Due to the overlap between the gate and ohmic contacts, the associated parallel parasitic capacitance must be de-embedded to access the CV characteristic solely between the gate and channel. This was achieved by extracting and deducting the saturated capacitance obtained when the channel was fully depleted for $V_{gs}$ values $> -6$ V i.e. sub-threshold. Integrating the de-embedded CV profile then allowed for extraction of hole density within the channel with respect to gate voltage, as shown in Fig. 9. The charge density vs gate bias profile follows a similar trend to the Transfer characteristics ($I_d$ vs $V_{gs}$) shown in Fig. 6, whereby a steady increase in charge density is observed with more negative $V_{gs}$ up to a maximum value approaching $2 \times 10^{12}$ cm$^{-2}$.

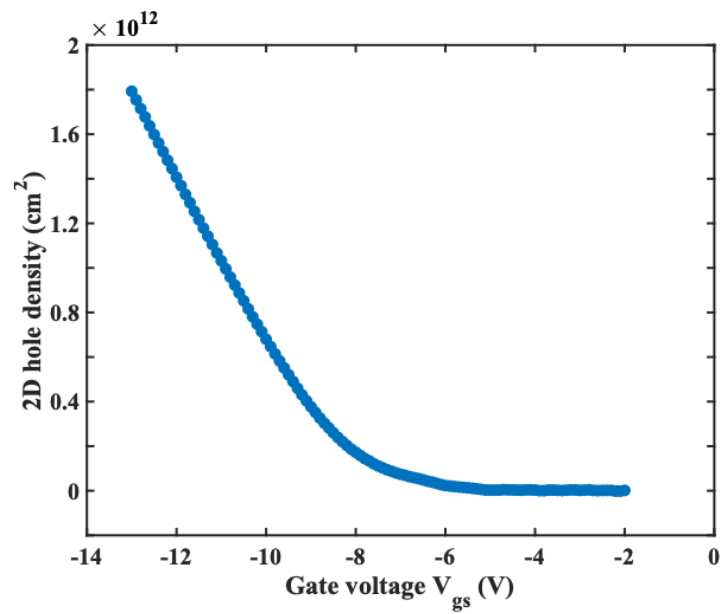

*Fig. 9 – Accumulation Channel 2D hole density vs $V_{gs}$ extracted by CV measurement from 1 μm gate/channel length, 25 μm wide Accumulation Channel FET.*

The channel sheet resistance, $R_{sh}$, was also extracted vs $V_{gs}$ by deducting the static total ohmic contact resistance deduced from TLM measurements from the device axial resistance ($R_{on}$) measured at various $V_{gs}$ values, as shown in Fig. 10.

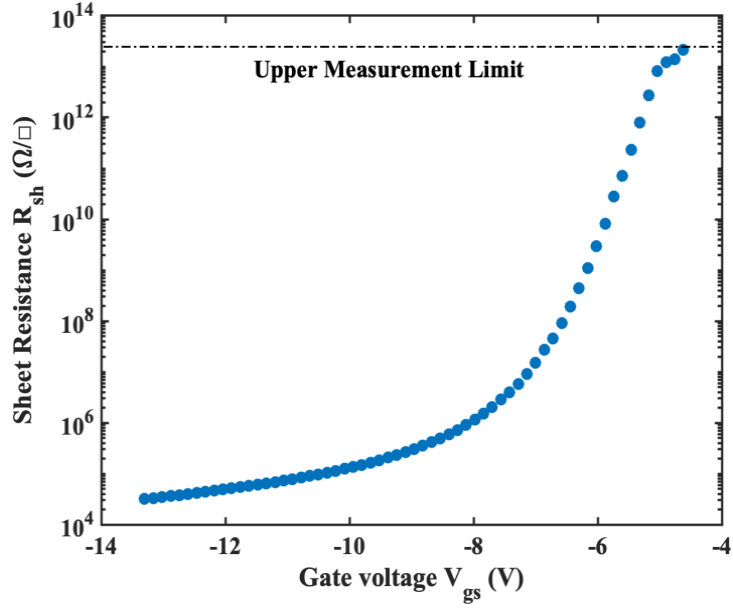

*Fig. 10 – Accumulation Channel $R_{sh}$ vs $V_{gs}$ calculated from total device axial resistance and ohmic contact resistance from 1 µm gate/channel length, 25 µm wide Accumulation Channel FET.*

By utilizing the following equation and combining hole density and $R_{sh}$ data presented in Fig 9. and Fig. 10, the effective mobility within the device channel is extracted vs $V_{gs}$ (Fig. 11):

$$\mu = \frac{1}{R_{sh}\, n\, q} \quad \text{[Eqn. 1]}$$

where $\mu$ is the average hole mobility, $R_{sh}$ the sheet resistance of the channel, $n$, the 2D hole density in the channel and $q$, the elemental value for charge.

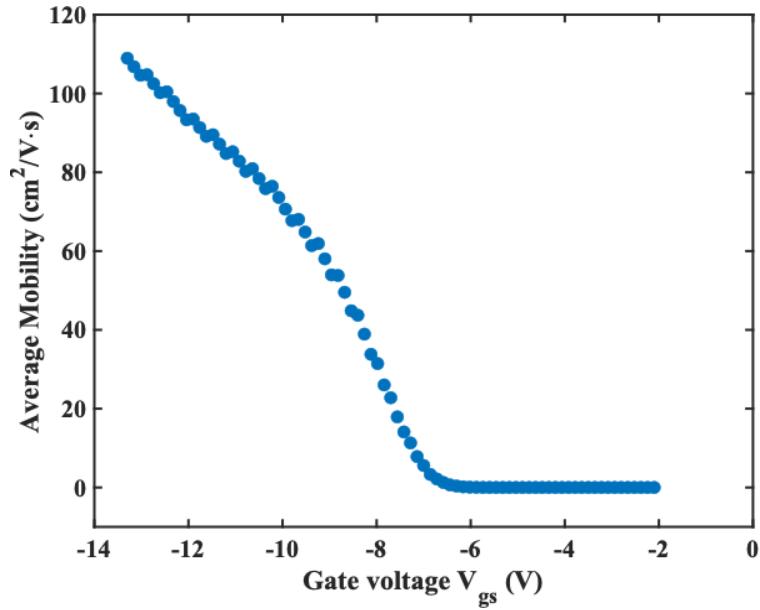

*Fig. 11 – Average hole mobility within the Accumulation Channel vs $V_{gs}$ extracted from 1 μm gate/channel length, 25 μm wide Accumulation Channel FET.*

Similar to the 2D hole density (Fig. 9), the hole mobility profile in the channel (Fig. 11) largely mirrors the trend of increasing $I_d$ with decreasing $V_{gs}$ observed in devices. A slight reduction in the rate of increase of hole mobility with decreased gate bias is observed at approximately $V_{gs}$ = - 9 V. Beyond this ($V_{gs}$ < -9 V), a steady albeit less steep mobility response is extracted until a maximum hole mobility ~ 110 cm²/V·s is observed at the most negative $V_{gs}$ inspected of -13V. The achieved average hole mobility in the device channel is comparable to previously reported values for H-terminated diamond in air with an equivalent carrier concentration of ~ $2 \times 10^{12}\ cm^{-2}$. However, this value is notably higher than mobilities reported for metal-oxide induced transfer-doping in H-diamond, which are typically around 60-70 cm²/V·s [37].

In contrast to lower hole mobility values reported for Transfer-Doped diamond devices, this improved mobility is most likely due to suppression of electron transfer to the contacting electron-beam deposited $Al_2O_3$ layer as associated with the Transfer Doping process and the resultant reduction in Coulomb scattering in the hole channel from this reduced trapped adjacent charge [22]. Several analyses of mobility limiting mechanisms for transfer-doped diamond [38–40] conclude that the primary scattering mechanisms are inhomogeneities in the channel potential, either due to inconsistencies in the hydrogen termination (termed C-H disorder by Peterson *et al.* [38]), or due to surface impurity scattering, as the separation between the electron acceptors at the surface and induced carrier concentration is reduced. The accumulation channel device architecture shown in the current work mitigates the latter, by removing the surface acceptors and associated random potential variations from the device channel via the 350°C pre-anneal prior to oxide deposition. This also has the impact of decoupling the active carrier concentration in the channel from the density of acceptors at the surface, i.e. the carrier concentration in the channel is dependent solely on the gate-bias, falling drastically in the subthreshold regime, as seen in Fig 9. Therefore, unlike in transfer-doped devices, in an accumulation channel device the increase in carrier concentration is no longer associated with increased surface-acceptor induced scattering, evident in Fig 12 where both average and differential mobility increases with increasing carrier concentration. This is further evidenced by the fact that substantial improvements in hole mobility have also been reported for H-diamond utilizing hexagonal boron nitride (hBN) as a surface dielectric layer to minimize scattering from near-interface trapped charge [15,40], which emphasizes the impact of a high quality, unpinned interface for performance enhancement in H-diamond FETs.

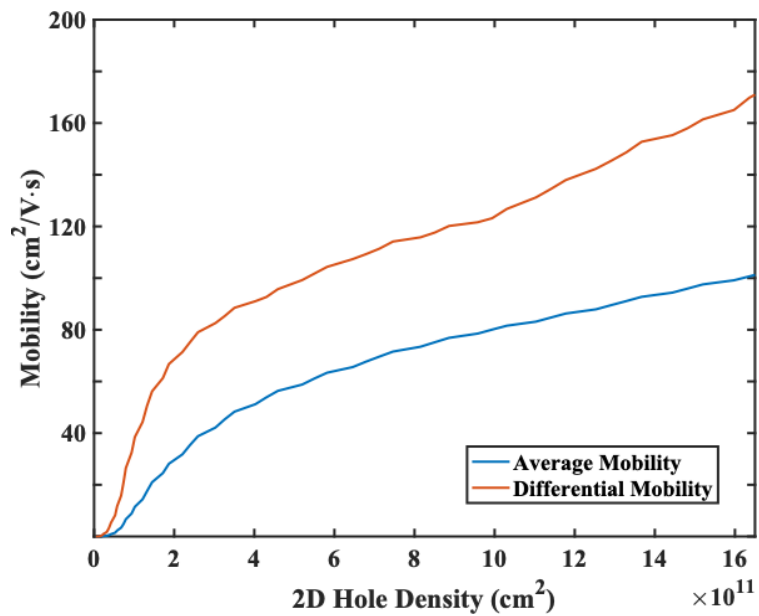

*Fig. 12 –Average and differential Hole Mobility vs 2D Hole density @ $V_{ds}$ = -10 V extracted from 1 μm gate/channel length, 25 μm wide Accumulation Channel FET.*

In Fig 12, the average hole mobility, as defined in Eqn. 1, has been plotted as a function of channel hole density. While the trend of mobility increasing with carrier density is contrary to the reciprocal relationship presented in Eqn 1., it is supported by percolation theory frameworks for mobility in the diamond 2DHG [38,41], described in detail in [3]. Under the percolation framework, devised initially for carrier transport in silicon surface inversion layers [42,43], a critical carrier concentration is required for carrier transport to exist through conductive 'percolation' channels, i.e. 2D states that extend the width of the quantum well, rather than phonon-assisted variable-range hopping that dominates at low carrier concentrations. This is explained as follows: when the carrier concentration in the channel is low and inhomogeneous, conductive regions are isolated from each other by insulating regions, and conduction occurs by thermal activation of carriers over the potential barriers that separate these regions. As the gate bias becomes increasingly negative, and correspondingly the hole density in the channel increases (see Fig 9), the 'percolation threshold' is crossed and allows for an increased mobility in the channel.

Furthermore, given the decoupling of the carrier density in the channel from the density of surface acceptors, Coloumb scattering from charged surface acceptors is no longer a consideration for mobility reduction with increasing carrier density – this is especially evident in the differential mobility, also known as the field-effect mobility, shown in Fig 12., which is defined by:

$$\mu_d = \mu + n\frac{d\mu}{dn} \qquad \text{[Eqn. 2]}$$

Where, $\mu$ is the average hole mobility as defined in Eqn. 1, and $n$ is the 2D hole density in the channel. The differential mobility is representative of the mobility of carriers added to the device channel with increasing gate-bias [44], and has been used in AlGaN/GaN HEMTs to more specifically assess the mobility of carriers in the on-state of the device [45]. In Fig 12, it is evident that the differential mobility increases rapidly with carrier concentration, which agrees with the premise that the carriers added to the channel via gate-field injection experience reduced scattering. This can be attributed both to the screening effect [46] with increasing carrier build-up in the channel, and remains in agreement with the percolation framework explained above, i.e. as the continuity of the 2DHG improves, the carriers are able to move more freely through the device channel. Critically, these results reaffirm the quality of the interface formed between the H-diamond surface and electron beam deposited $Al_2O_3$, with which both channel carrier density and mobility maintain an increasing trend with increased proximity of charge to the interface.

Inspection of the intrinsic operation of devices was also performed to examine the impact of the source contact resistance ($R_c$) upon transconductance. Using a value of 11.4 Ω·mm, for $R_c$ (as determined by TLM measurement), intrinsic transconductance ($g_m$*) is extracted using the

following expression:

$$\text{Intrinsic transconductance, } g_m* = \frac{g_m}{1-(R_c\, g_m)} \quad \text{[Eqn. 3.]}$$

Extrinsic and intrinsic transconductance ($g_m$ & $g_m*$ vs $V_{gs}$ @ $V_{ds}$ = -10 V) curves are plotted for comparison in Fig. 13:

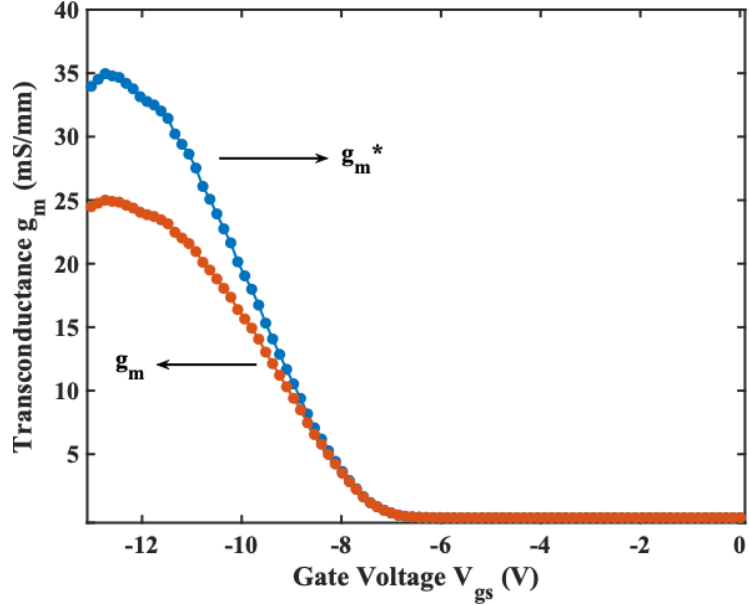

*Fig. 13 – Extrinsic transconductance ($g_m$) & Intrinsic transconductance ($g_m*$) vs $V_{gs}$ @ $V_{ds}$ = -10 V extracted from 1 μm gate/channel length, 25 μm wide Accumulation Channel FET.*

De-embedding the source contact resistance produces a peak intrinsic transconductance value of 35 mS/mm, corresponding to a 40% increase compared to peak extrinsic transconductance. Extraction of the intrinsic transconductance additionally allowed for determination of the average hole velocity in the device channel under peak transconductance bias conditions using the following expression:

$$\text{Average channel hole velocity, } v_h = \frac{g_m*\, L_{ch}}{C_g} \quad \text{[Eqn. 3]}$$

Where: $g_m*$ is the intrinsic transconductance, $L_{ch}$ the channel length (i.e. 1 μm) and $C_g$ the gate capacitance. Utilizing this relationship, a peak value for the average hole velocity in the channel of $5.62 \times 10^6\, cm/s$ was extracted across the inspected gate bias range. It should be noted that this most likely represents an underestimated velocity figure, due to the gate capacitance used in this calculation being extracted at $V_{ds}$ = 0V, rather than at more negatively $V_{ds}$ i.e. the associated reduction in carrier density towards the drain end of the channel under non-zero $V_{ds}$ will reduce the total $C_g$ for a given $V_{gs}$, leading to higher hole velocity as per Eqn. 3. This value however is in good agreement with hole velocity values reported from H-diamond transfer-doped FETs [47].

The very negative threshold voltage achieved in the devices presented in this work is again attributed to the implementation of an Accumulation Channel rather than Transfer Doped device architecture, whereby spontaneous formation of the 2D hole channel is suppressed until a significantly negative gate voltage is applied. These results are compared with the state of the art (to the best of our knowledge) E-mode H-diamond FETs reported in literature to date, shown in Fig. 14.

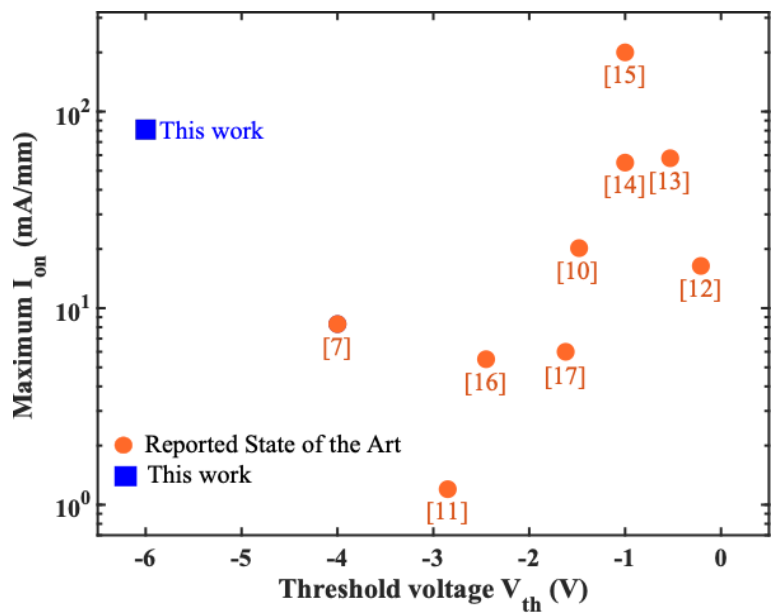

*Fig. 14 – Comparison of state-of-the-art E-mode H-diamond FETs reported to date, showing device maximum on-current ($I_{on}$) vs extracted threshold voltage.*

As shown in Fig. 14, the majority of E-mode H-diamond FETs reported demonstrate threshold voltage typically higher (i.e. more positive) than -3 V, and often with lower on-current than that achieved in D-mode H-diamond FETs. A notable exception is [7], whereby a more negative $V_{th}$ of ~ -4V is achieved through partial oxygen-termination of the H-diamond surface to deliberately reduce carrier density within the hole channel. In contrast to the work presented here, whereby a H-terminated surface and hence unpinned Fermi level is maintained, this mixed termination most likely leads to Fermi pinning [48–50] and hence limits the ability to accumulate significant charge density in the channel for high $I_{on}$. It should be noted again that the often-differing techniques used to extract $V_{th}$ across these studies may lead to more pessimistic or optimistic threshold voltage values depending on the technique used in question (as we have demonstrated in the $V_{th}$ analysis in this paper), which makes direct and fair comparison in this summary challenging. As such readers are advised to refer to the various refences in question for further investigation into this summarized data. It should be noted again however that in the devices presented here, the drain current is below the measurement system noise floor for $V_{gs} > 5$ V, demonstrating an indisputable $V_{th}$ of at least -5 V for this technology and the lowest yet reported for H-diamond FETs. Furthermore, other commonly used threshold voltage extraction techniques when applied to these devices demonstrated unparalleled values as low as – 9 V. In contrast to these results for H-terminated diamond-based FETs, recent work reported on Si-terminated diamond FETs has also demonstrated extremely low $V_{th}$ values down to -7 V with similar $I_{on}$ (~100 mA/mm)[51,52], signifying interesting potential to develop high performance diamond transistors with non-traditional surface chemistries.

5. Conclusion

Accumulation Channel H-diamond FETs are reported which demonstrate the lowest threshold voltage yet achieved for H-diamond FET technology, while also delivering competitively high on-current. This is achieved through engineered inhibition of the Transfer Doping process to induce extremely high off-state resistance and reduce Coulomb scattering to increase hole mobility while maintaining a high-quality H-diamond interface to maximize charge accumulation potential for high on-current. Results indicate that even further improvement to $I_{on}$ should also be achievable through optimization of both the ohmic contact formation process and dielectric materials strategy used for transfer-doping suppression. Furthermore, the relatively low off-state breakdown observed in the reported devices highlights the deficiency in the presented FET structure for high power applications i.e. the close proximity of the gate contact to the ohmic contacts. The techniques presented here to achieve record normally-off operation should however find application in future hybrid diamond FETs which adopt both Accumulation Channel and Transfer Doping approaches in a single device architecture to simultaneously achieve high on-current, high off-state breakdown and enhanced E-mode operation.

Acknowledgment

The authors wish to acknowledge the staff and facilities of the James Watt Nanofabrication Centre.